\documentclass[a4paper,11pt]{article}
\pdfoutput=1

\usepackage{array}
\usepackage{amsfonts}
\usepackage{amsmath,amssymb}
\usepackage{cite}
\usepackage{color}
\usepackage[top=1.2in, bottom=1.2in, left=1.1in, right=1.1in]{geometry}
\usepackage{graphicx,subfigure}
\graphicspath{{./}{./figures/}}
\usepackage{mathrsfs} 
\usepackage{pifont}
\usepackage[small]{caption}
\usepackage{multirow}
\usepackage{slashed}
\usepackage{ulem}

\definecolor{mediumblue}{rgb}{0,0,0.8}
\usepackage{hyperref}
\hypersetup{
	linktocpage=true,
	colorlinks=true,
	citecolor=mediumblue,
	filecolor=mediumblue,
	linkcolor=mediumblue,
	urlcolor=mediumblue,
}
\newcommand{\mailref}[1]{\href{mailto:#1}{#1}}

\allowdisplaybreaks

\def\beq{\begin{equation}}
\def\eeq{\end{equation}}
\def\beqr{\begin{eqnarray}}
\def\eeqr{\end{eqnarray}}
\def\bdpm{\begin{displaymath}}
\def\edpm{\end{displaymath}}
\def\half{\frac{1}{2}}
\definecolor{lgray}{gray}{0.6}

\newcommand{\nnb}{\nonumber}

\newcommand{\tbt}{t_\beta}
\newcommand{\tth}{t_\theta}

\newcommand{\lamh}{\lambda_h}
\newcommand{\lams}{\lambda_s}
\newcommand{\lamp}{\lambda_\phi}
\newcommand{\lamhs}{\lambda_{hs}}
\newcommand{\lamhp}{\lambda_{h\phi}}
\newcommand{\lamsp}{\lambda_{s\phi}}

\begin{document}

\begin{titlepage}
\def\thefootnote{\fnsymbol{footnote}}
	
\begin{flushright}
	\texttt{}\\
	\texttt{}
\end{flushright}
	
\begin{centering}
\vspace{0.5cm}
{\Large \bf \boldmath
	Scalar dark matter multiplet of global $O(N)$ symmetry }

\bigskip

\begin{center}
	{\normalsize 
		U-Rae Kim$^{a,}$\footnote{\mailref{kim87@kma.ac.kr}},
		Jungil Lee$^{b,}$\footnote{\mailref{jungil@korea.ac.kr}}, and
		Soo-hyeon~Nam$^{b,}$\footnote{\mailref{glvnsh@gmail.com, corresponding author}}}\\[0.5cm]
	\small
	$^a${\em Department of Physics, Korea Military Academy, Seoul 01805, Korea}\\[0.1cm]
	$^b${\em Department of Physics, Korea University, Seoul 02841, Korea}

\end{center}
\end{centering}

\medskip

\begin{abstract}
   \noindent
We study two types of models involving a scalar dark matter multiplet of global $O(N)$ symmetry. 
These models are distinguished by the absence (Type I) or presence (Type II) of a scalar mediator with $Z_{2}$ symmetry. 
We derive the allowed regions for the dark matter mass and new scalar couplings based on constraints 
from Higgs invisible decay, the relic abundance of dark matter, 
and the spin-independent dark matter-nucleon scattering cross section. 
Within the allowed parameter space, we also discuss the vacuum stability of the Higgs potential 
and the perturbativity of the scalar couplings in both models. 
We find that the Type I model cannot achieve stable electroweak vacuum, 
whereas the Type II model can have both a stable vacuum and perturbative couplings up to the Planck scale. 
\end{abstract}

\vspace{0.5cm}

\end{titlepage}

\renewcommand{\thefootnote}{\arabic{footnote}}
\setcounter{footnote}{0}

\setcounter{tocdepth}{2}
\noindent \rule{\textwidth}{0.3pt}\vspace{-0.4cm}\tableofcontents
\noindent \rule{\textwidth}{0.3pt}

\section{Introduction\label{sec:intro}}

The standard model (SM) of particle physics has been very successful in describing the known fundamental particles and their interactions. 
However, it does not account for non-baryonic cold dark matter (DM) in the universe. 
The existence of DM is strongly supported by various astrophysical observations 
such as galaxy rotation curves, gravitational lensing, and cosmic microwave background anisotropies. 
Since the particle nature of DM remains unknown, numerous extensions to the SM have been proposed to identify a viable DM candidate, 
often included in a separate hidden sector connected to the SM. 
In particular, scalar DM has been widely studied because it can directly couple to the Higgs scalar and potentially alter Higgs phenomenology 
\cite{Silveira:1985rk,McDonald:1993ex,Holz:2001cb,Burgess:2000yq,He:2007tt,He:2011de,Drozd:2011pex}. 

One significant advantage of coupling a new scalar field to Higgs is that 
it allows modification of the renormalization group running of the Higgs quartic coupling. 
In the SM, quantum corrections to the Higgs quartic coupling could drive it to negative values at high energies, 
leading to the electroweak vacuum becoming metastable \cite{Isidori:2001bm,Degrassi:2012ry,Alekhin:2012py,Buttazzo:2013uya,Bednyakov:2015sca}. 
This vacuum instability in the Higgs potential implies that the universe could make a transition to a lower energy state, 
contradicting the long-term stability we currently observe. 
However, this issue can be  resolved by introducing direct interactions between a scalar DM and Higgs 
\cite{Gonderinger:2009jp,Lebedev:2012zw}.

In this work, we explore DM models that include hidden sector DM scalar particles 
with a global $O(N)$ symmetry imposed to ensure the stability of the hidden sector.
This study includes two models distinguished by the absence (Type I) or presence (Type II) of a scalar mediator.
In the Type I model, the nonzero SM-DM coupling ($\lamhp$) plays a crucial role between the SM and the hidden sector interactions.
As a result, it is tightly constrained by both collider experiments and DM phenomenological observables. 
Conversely, the Type II model introduces a scalar mediator which modifies the interaction landscape considerably.
This scenario relaxes constraints on the coupling $\lamhp$ 
while facilitating a complex interplay between the Higgs sector and the hidden sector, 
offering various degrees of freedom to explore the resultant phenomenology 
and its implications for the Higgs boson and vacuum stability.
Taking into account all theoretical considerations,
we derive conservative bounds on the coupling $\lamhp$ and the DM mass, 
and identify parameter sets with stable vacuum that satisfy the current phenomenological constraints.

This paper is organized as follows. In Sec.~II, we describe the Type I and II models in detail. 
Next, we provide a detailed phenomenological analysis of the DM physics such as the relic density and the direct detection
as well as of the collider physics in Sec.~III. 
In Sec.~IV, we show the $\beta$-functions of the couplings and discuss the vacuum stability and perturbativity conditions.
Section V summaries the results and concludes.

\section{Model}

\subsection{Type I: DM scalar multiplet}

We consider a simple model of a dark sector consisting of a real SM gauge singlet scalar field $\phi$, 
which is a DM candidate chosen to be the fundamental representation of a global $O(N)$ group, 
$\phi = (\phi_1, \cdots, \phi_N)^T$. 
The extended Higgs sector Lagrangian with the renormalizable DM interactions is then given by
\beq \label{eq:DM_Lagrangian}
\mathscr{L}_\textrm{I}^{\rm DM} = \left(D_\mu H\right)^{\dagger}D^\mu H 
+\half(\partial_\mu\phi^T)\partial^\mu\phi - V_\textrm{I}(H, \phi), 
\eeq
where the scalar potential is
\beq \label{eq:Vphi_potential}
V_\textrm{I}(H, \phi) = \mu_H^2 H^{\dagger} H + \lamh (H^{\dagger} H)^2 
+ \half \mu_\phi^2\phi_i^2 + \half\lambda_{h\phi} H^{\dagger} H \phi_i^2 
+ \frac{1}{4}\lambda_\phi (\phi_i^2)^2,
\eeq
with $\phi_i^2 = \phi^T\phi$.
This model was previously studied in Ref.~\cite{Drozd:2011pex}, 
considering the low DM mass region of less than 140 GeV while focusing on the sub-1 GeV range.
However, we study the entire range of the DM mass.
If $N=1$, this model turns into the usual scalar ``Higgs-portal" scenario with a $Z_2$ symmetry 
\cite{Silveira:1985rk,McDonald:1993ex,Holz:2001cb,Burgess:2000yq,He:2007tt,He:2011de}.
The DM scalar $\phi$ does not obtain a VEV due to the unbroken $O(N)$ symmetry,
and this model has only three free parameters: $\mu_\phi$, $\lambda_{h\phi}$, and $\lambda_{\phi}$.

Following the electroweak symmetry breaking (EWSB), 
The tree-level scalar potential can be expressed in the unitary gauge as
\beq \label{eq:Vphi_interaction}
V_\textrm{I}(h, \phi_i) = \lamh v_h^2 h^2 + \lamh v_h h^3 + \frac{\lamh}{4}h^4 
+ \half M_{\phi} \phi_i^2 + \frac{\lamhp}{4}\left(h^2 + 2v_h h\right)\phi_i^2 + \frac{\lamp}{4}(\phi_i^2)^2,
\eeq
where $h$ is the physical Higgs boson, $v_h$ is the VEV of the Higgs field,
and $M_{\phi}$ is the physical mass of the DM scalar $\phi$ given by
\beq
M_{\phi}^2 = \mu_\phi^2 + \frac{v_h^2}{2}\lambda_{h\phi}.
\eeq
Dark matter phenomenology is predominantly governed by 
two independent parameters $M_\phi$ and $\lambda_{h\phi}$.
The DM self-coupling $\lambda_{\phi}$ is irrelevant to the SM-DM interactions,
but it affects the DM self-interactions \cite{Bento:2000ah,McDonald:2001vt,Bernal:2015xba,Tulin:2017ara} 
and the renormalization group equation (RGE) of the other couplings considerably \cite{Kim:2022sfc}.

\subsection{Type II: DM scalar multiplet + scalar mediator}

In addition to the scalar DM $\phi$, 
we adopt a scalar mediator $S$ which is responsible for the EWSB together with the SM Higgs doublet $H$.
The extended Higgs sector Lagrangian is then modified as
\beq \label{eq:DM_Lagrangian}
\mathscr{L}^{\rm DM}_\textrm{II} = \left(D_\mu H\right)^{\dagger}D^\mu H 
+ \half\left(\partial^{\mu} S\right)^2 +\half(\partial_\mu\phi^T)\partial^\mu\phi - V_{\rm II}(H, S, \phi), 
\eeq
where the scalar potential is extended as
\beq \label{eq:VS_potential}
V_\textrm{II}(H, S, \phi) = V_\textrm{I}(H, \phi) + \half \mu_S^2 S^2 
+ \half\lamhs H^{\dagger} H S^2 + \frac{1}{4}\lams S^4 + \frac{1}{4}\lambda_{s\phi} S^2\phi_i^2.
\eeq
The classically scale-invariant version of this model was studied in our previous work \cite{Jung:2019dog}, 
and we impose $Z_2$ symmetry on $S$ for clear comparison with the scale-invariant case.
For $N = 1$ case, a similar model was studied in Ref.~\cite{Claude:2021sye},
but they simply assumed $\lamhp = 0$ in order to decouple the DM sector from the SM Higgs. 
In general, however, such an interaction term is not forbidden by a
discrete symmetry such as $Z_2$ symmetry theoretically, and also it is very important to explain the current astronomical
observables phenomenologically and the vacuum stability of the scalar potential, 
as discussed in our earlier studies \cite{Jung:2019dog,Kim:2022sfc}.

After the EWSB, the singlet scalar $S$ also develop nonzero VEV, $\langle S \rangle = v_s$.
The potential minimization conditions
$\partial V/\partial H |_{\langle H^0\rangle=v_h/\sqrt{2}} = \partial V/\partial S |_{\langle S\rangle=v_s}=0$
lead to the following relations:
\beq \label{eq:lambdamin}
\mu_H^2 = -\lamh v_h^2 - \half \lamhs v_s^2 , \qquad
\mu_S^2 = -\lams v_s^2 - \half \lamhs v_h^2.
\eeq
The neutral scalar fields $h$ and $s$ defined by $H^0=(v_h+h)/\sqrt{2}$ and $S=v_s+s$ 
are mixed to yield the mass matrix:
\beq \label{eq:mass_mattrix}
M^2 = \left( \begin{array}{cc} 
2 \lamh v_h^2 &\ \lamhs v_h v_s
\\[1pt]
\lamhs v_h v_s &\ 2 \lams v_s^2 
\end{array} \right) .
\eeq
In terms of the gauge eigenstates, the tree-level scalar potential can be expressed 
in the unitary gauge as
\beqr \label{eq:VS_interaction}
V_\textrm{II}(h, s, \phi_i) &=& V_\textrm{I}(h, \phi_i) + \lams v_s^2 s^2 + \lams v_s s^3 + \frac{\lams}{4}s^4 \nnb \\[1pt]
&& +\lamhs \left( v_h v_s h s + \half v_s h^2 s + \half v_h h s^2 + \frac{1}{4} h^2 s^2 \right) 
+ \frac{\lamsp}{4}\left(s^2 + 2v_s s\right)\phi_i^2.
\eeqr
The corresponding scalar mass eigenstates $h_1$ and $h_2$ are admixtures of $h$ and $s$:
\beq
\left( \begin{array}{c} h_1 \\[1pt] h_2 \end{array} \right) =
\left( \begin{array}{cc} \cos \theta &\ \sin \theta \\[1pt]
-\sin \theta &\ \cos \theta \end{array} \right)
\left( \begin{array}{c} h \\[1pt] s \end{array} \right) ,
\eeq
where the mixing angle $\theta$ is given by
\beq \label{eq:tan_theta}
\tan \theta 
= - \frac{\lamh-\lams\tbt^2 -\sqrt{(\lamh-\lams\tbt^2)^2+\lamhs^2\tbt^2}}
{\lamhs\tbt} 
\eeq
and where $\tbt\, (\equiv \tan\beta) = v_s/v_h$.
The mixing angle $\theta$ is expected to be very small (less than about 0.25) due
to the LEP constraints \cite{LEPWorkingGroupforHiggsbosonsearches:2003ing}.
After diagonalizing the mass matrix, 
we obtain the physical masses of the two scalar bosons ($h_1, h_2$) and the DM scalar $\phi$ as follows:
\beq
M^2_1 = \frac{2v_h^2(\lamh-\lams\tbt^2\tth^2)}{1-\tth^2}, \quad
M^2_2 = \frac{2v_h^2(\lams\tbt^2-\lamh\tth^2)}{1-\tth^2}, \quad
M_{\phi}^2 = \mu_\phi^2 + \frac{v_h^2}{2}\left(\lambda_{h\phi} + \lambda_{s\phi}\tbt^2\right),
\eeq
where $\tth \equiv \tan\theta$.
We assume that $M_1$ corresponds to the observed SM-like Higgs boson mass in what follows.
In terms of the physical states, the tree-level scalar potential can be expressed 
in the unitary gauge as
\beqr \label{eq:V_interaction}
V_\textrm{II}(h_1, h_2, \phi_i) &=& \half M_1^2 h_1^2 + \half M_2^2 h_2^2 + \half M_\phi^2 \phi_i^2 \nnb \\[1pt]
&& +\frac{1}{4(1+\tth^2)^2}\left[\lamh+\lams\tth^4 
+\left(\frac{\lamh}{\tbt}-\lams\tbt\right)\frac{2\tth^3}{1-\tth^2}\right]h_1^4 \nnb \\[1pt]
&& +\frac{1}{4(1+\tth^2)^2}\left[\lamh\tth^4+\lams 
+\left(\frac{\lamh}{\tbt}-\lams\tbt\right)\frac{2\tth^3}{1-\tth^2}\right]h_2^4 \nnb \\[1pt]
&& +\frac{\tth}{2(1+\tth^2)^2}\left[\lamh\left(3\tth+\frac{1-4\tth^2+\tth^4}{(1-\tth^2)\tbt}\right)
+\lams\left(3\tth-\frac{1-4\tth^2+\tth^4}{1-\tth^2}\tbt\right)\right]h_1^2h_2^2 \nnb \\[1pt]
&& +\frac{\tth(\tth-\tbt)(\lamh+\lams\tth\tbt)}{(1+\tth^2)^2\tbt}h_1^3h_2
-\frac{\tth(1+\tth\tbt)(\lamh\tth-\lams\tbt)}{(1+\tth^2)^2\tbt}h_1h_2^3 \nnb \\[1pt] 
&& +\frac{(\tth^3+\tbt)(\lamh-\lams\tth^2\tbt^2)}{(1-\tth^2)(1+\tth^2)^{\frac{3}{2}}\tbt}v_hh_1^3
-\frac{(1-\tth^3\tbt)(\lamh\tth^2-\lams\tbt^2)}{(1-\tth^2)(1+\tth^2)^{\frac{3}{2}}\tbt}v_hh_2^3 
\nnb \\[1pt] 
&& +\frac{\tth(\tth-\tbt)\left(\lamh (2-\tth^2)+\lams (1-2\tth^2)\tbt^2 \right)}
{(1-\tth^2)(1+\tth^2)^{\frac{3}{2}}\tbt}v_hh_1^2h_2 \nnb \\[1pt] 
&& +\frac{\tth(1+\tth\tbt)\left(\lamh (1-2\tth^2)+\lams (2-\tth^2)\tbt^2 \right)}
{(1-\tth^2)(1+\tth^2)^{\frac{3}{2}}\tbt}v_hh_1 h_2^2 \nnb \\[1pt] 
&& +\frac{\lambda_{h\phi}+\lambda_{s\phi}\tth^2}{4(1+\tth^2)}h_1^2\phi_i^2
+\frac{\lambda_{h\phi}\tth^2+\lambda_{s\phi}}{4(1+\tth^2)}h_2^2\phi_i^2
-\frac{\tth(\lambda_{h\phi}-\lambda_{s\phi})}{2(1+\tth^2)}h_1h_2\phi_i^2 \nnb \\[1pt]
&& +\frac{\lambda_{h\phi}+\lambda_{s\phi}\tth\tbt}{2\sqrt{1+\tth^2}}v_h h_1\phi_i^2
-\frac{\lambda_{h\phi}\tth-\lambda_{s\phi}\tbt}{2\sqrt{1+\tth^2}}v_h h_2\phi_i^2
+\frac{1}{4}\lambda_\phi (\phi_i^2)^2.
\eeqr
Note that we replaced $\lamhs$ with $\tth$ using the relations in Eq.~(\ref{eq:tan_theta}).
The dependencies of the model parameters are 
\beq \label{eq:couplings}
v_s = v_h\tbt, \quad
\lamh = \frac{M_1^2 + M_2^2\tth^2}{2v_h^2(1+\tth^2)}, \quad
\lams = \frac{M_1^2\tth^2 + M_2^2}{2v_h^2\tbt^2(1+\tth^2)}, \quad
\lamhs = \frac{(M_1^2 - M_2^2)\tth}{v_h^2\tbt(1 + \tth^2)} .
\eeq
It is clear from the above equation that $\tbt$ should not be very small 
because of the perturbativity of the couplings $\lambda_{hs,s}$.
Given the fixed Higgs mass $M_1$ and $v_h \simeq 246$ GeV, 
there are seven independent new physics (NP) parameters:
$M_2, \tbt, \tth, M_\phi, \lambda_{h\phi}, \lambda_{s\phi}, \lambda_\phi $.
We constrain these NP parameters by taking into account various theoretical considerations and
experimental measurements in the following sections.

\section{Phenomenology}

We consider phenomenological implications and interactions of new particles in our models at colliders,
and discuss DM phenomenology in this section. 

\subsection{Type I}

Since the hidden sector is connected to the SM by the Higgs portal in this model, 
there are constraints from the collider experiments.
Due to the nonzero SM-DM coupling ($\lamhp$), if $M_\phi < M_h/2$, 
the SM Higgs $h$ can decay invisibly into a pair of DM through mixing with decay width,
\beq
\Gamma_\textrm{inv.} = \Gamma\left(h \to \phi_i\phi_i\right)
= \frac{\lamhp^2 v_h^2}{32\pi M_h}\sqrt{1-\frac{4M_\phi^2}{M_h^2}} ,
\eeq
and the corresponding branching fraction of the invisible Higgs decays is given by the relation
BR($h \to$ inv.) = $\Gamma_\textrm{inv.}/(\Gamma_\textrm{SM} + \Gamma_\textrm{inv.})$,
where $\Gamma_\textrm{SM}$ = 4.07 MeV \cite{LHCHiggsCrossSectionWorkingGroup:2016ypw}.
The most recent upper limit on the Higgs invisible decay has been set 
by the ATLAS Collaboration using the luminosity of 139 fb$^{-1}$ data 
at center-of-mass energy of 13 TeV recorded in Run 2 of the LHC \cite{ATLAS:2023tkt}.
We applied the combined 95\% confidence-level limit of BR($h \to$ inv.) $<$ 0.107 to our model.
Beside the invisible Higgs decay bound applied in low DM mass region, 
an upper bound of the DM mass can be also obtained 
from the partial-wave unitarity of the $S$-matrix combined with the updated relic density.
Following Ref.~\cite{Griest:1989wd}, 
we obtain a naive upper bound on $M_\phi\sim 117$ TeV for our models.
As we will see soon below, however, our results are not constrained by the above bound. 

Next, we consider the relic density constraint on this model. 
The current determination of the DM mass density $\Omega_{\rm DM}$ 
comes from global fits of cosmological parameters to a variety of observations
such as Planck primary cosmic microwave background (CMB) data plus the Planck measurement of CMB lensing 
\cite{Planck:2018vyg}:
\beq
\Omega_\textrm{CDM}h^2 = 0.1200 \pm 0.0012.
\label{eq:relic_obserb}
\eeq
This relic density observation will exclude some regions in the model parameter space.
The relic density analysis in this section includes all possible channels of $\phi_i\phi_i$ pair annihilation into the SM particles. 
Using the numerical package micrOMEGAs \cite{Belanger:2018ccd}
that utilizes CalcHEP for computing the relevant annihilation cross sections \cite{Belyaev:2012qa}, 
we compute the DM relic density and the spin-independent (SI) DM-nucleon scattering cross sections.

\begin{figure}[!hbt]
\centering%
\subfigure[ Type I ]{\label{fig:sigmaSI1} %
\includegraphics[width=7.4cm]{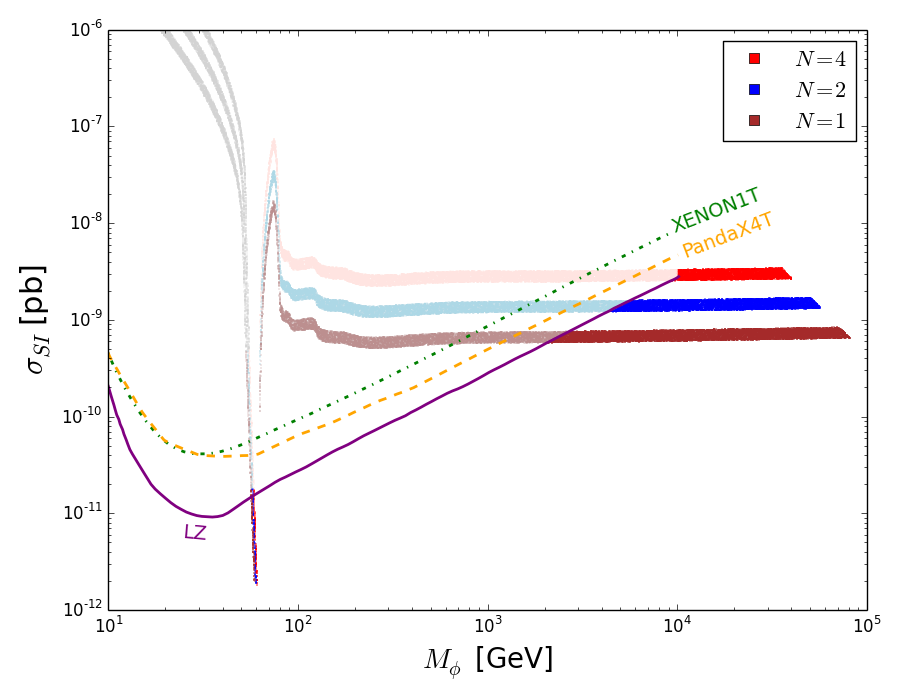}} \
\subfigure[ Type II ]{\label{fig:sigmaSI2} %
\includegraphics[width=7.4cm]{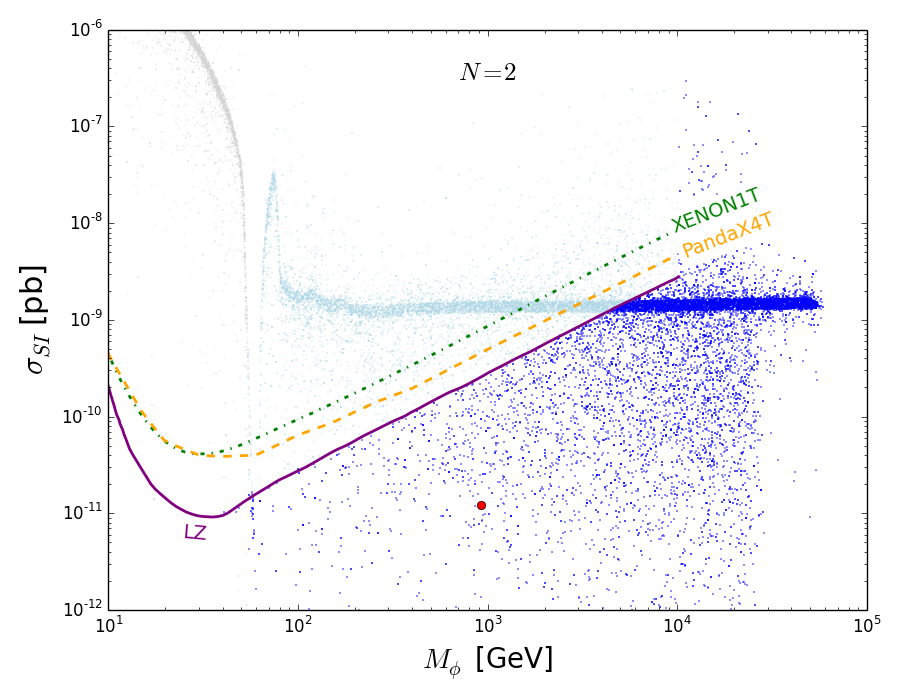}} 
\caption{Spin-independent DM-nucleon scattering results allowed by relic density observations. 
Also shown are observed limits from XENON1T \cite{XENON:2018voc}, PandaX4T \cite{PandaX-4T:2021bab}, 
and LUX-ZEPLIN (LZ) \cite{LZ:2022lsv}.
Some of allowed parameter sets by the relic density observation in low DM mass (grayed) region 
are excluded by the invisible Higgs decay bound, 
and the excluded region by the LZ bound is plotted in lighter colors.
The red dot in the right figure indicates a benchmark point chosen to discuss the running behavior
of the scalar couplings.
} 
\label{fig:sigmaSI}
\end{figure}

For illustration of allowed new model parameter spaces, 
we fix the DM self-coupling $\lambda_{\phi} = 0.01$ and consider the $N=1, 2,$ and 4 cases. 
A large value of $N$ is disfavored because this ruins perturbativity of the scalar couplings at high scales.
Using the relic density constraint given in Eq.~(\ref{eq:relic_obserb}), 
we perform the phenomenological analysis of the model by varying the following 
two NP parameters: $M_{\phi}$, $\lambda_{h\phi}$. 
In Fig. \ref{fig:sigmaSI1}, 
we plot the SI DM-nucleon scattering cross section 
by varying the DM mass $M_\phi$ with parameter sets allowed by the relic density observation within $3\sigma$ range
and compare the results with the observed upper limits obtained at the 90$\%$ confidence level 
from XENON1T \cite{XENON:2018voc}, PandaX4T \cite{PandaX-4T:2021bab}, and LUX-ZEPLIN (LZ) \cite{LZ:2022lsv}.
Some of newer observation results are not shown since those are given below 1 TeV
and not much different from the above results.
In low DM mass region, the grayed data are excluded by the invisible Higgs decay bound. 
Nonobservation of DM-nucleon scattering events is interpreted as an upper bound on the DM-nucleon cross section.
The data excluded by the LZ bound are plotted in lighter colors on the plot. 
DM-nucleon scattering occurs via a $t$-channel diagram exchanging the Higgs boson, 
leading to resonance effects near half of the Higgs mass.
The figure also shows that, aside from the resonance region, 
larger values of $N$ impose stronger constraints. 
The lower bounds on the DM masses are approximately 
2.1 TeV, 4.4 TeV, and 9.4 TeV for $N$ = 1, 2, and 4, respectively. 
We also investigated larger $N$ cases and found that for $N \geq 11$, the scenarios are ruled out by the LZ bound
when extrapolated beyond 10 TeV.

\begin{figure}[!hbt]
\centering%
\subfigure[ Type I ]{\label{fig:lamhp_vs_Mdm1} %
\includegraphics[width=7.4cm]{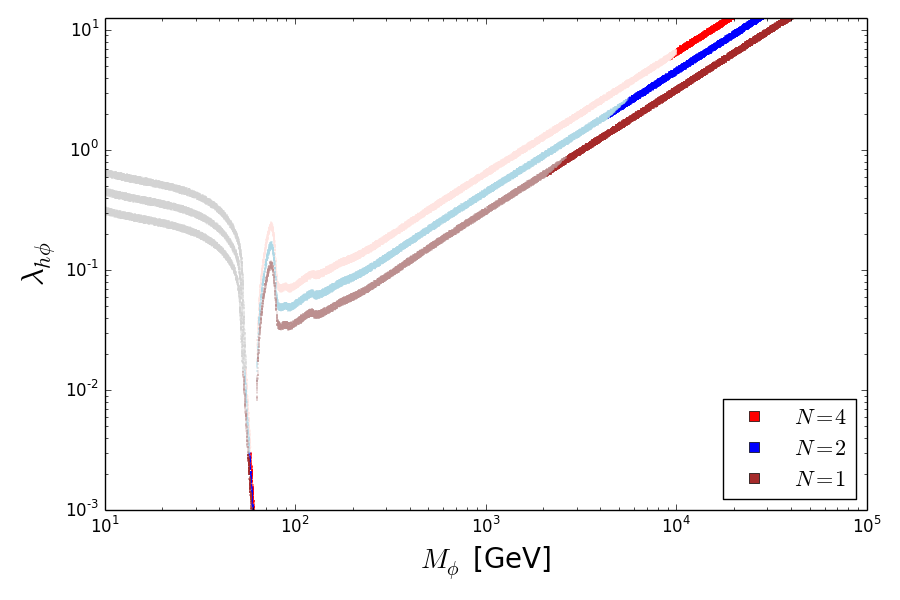}} \
\subfigure[ Type II ]{\label{fig:lamhp_vs_Mdm2} %
\includegraphics[width=7.4cm]{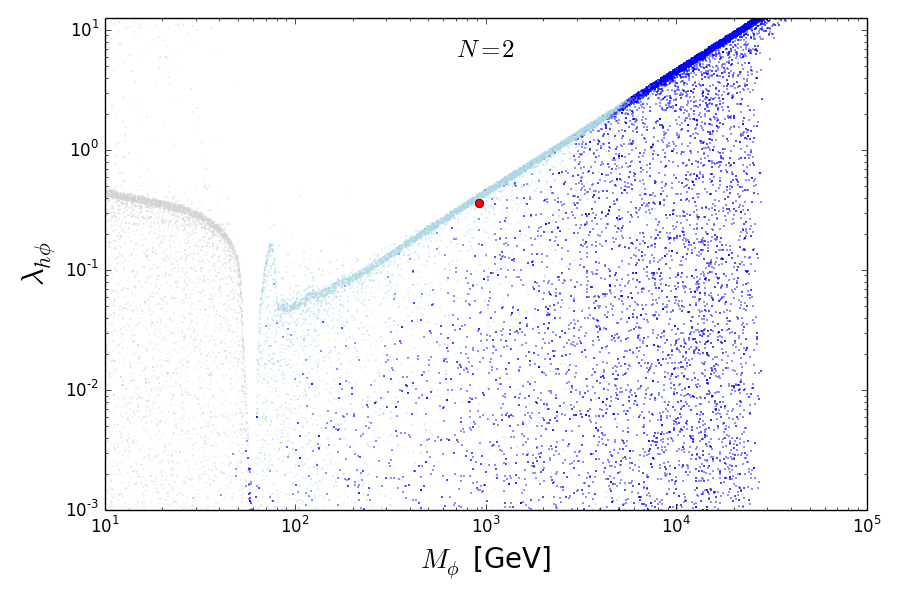}} 
\caption{Allowed regions for the parameter sets of ($M_\phi, \lambda_{h\phi}$) 
by relic density observations at 3$\sigma$ level. 
Some of allowed parameter sets by the relic density observation in low DM mass (grayed) region 
are excluded by the invisible Higgs decay bound,
and the excluded region by the direct detection bound from LZ experiment is plotted in lighter colors.
The red dot in the right figure indicates a benchmark point chosen to discuss the running behavior
of the scalar couplings. 
} 
\label{fig:lamhp_vs_Mdm}
\end{figure}

To see the relic density constraints on the scalar DM interaction,
we plot the allowed region of the Higgs-DM coupling $\lambda_{h\phi}$ versus the DM mass $M_\phi$
constrained by current relic density observations at 3$\sigma$ level in Fig.~\ref{fig:lamhp_vs_Mdm1}.
The figure indicates that $\lambda_{h\phi}$ must be either less than approximately 0.003 
or greater than about 0.6 (for $N=1$), 1.8 (for $N=2$), or 2.7 (for $N=4$)
to satisfy all phenomenological constraints. 
We will discuss the running behavior of the couplings for $\lambda_{h\phi}(v_h) = 0.6$ and 1.8 cases
in the next section.

\subsection{Type II}

Due to the Higgs-portal terms in Eq.~(\ref{eq:VS_potential}), 
the electroweak interaction of the Higgs boson can be significantly modified, 
and it is possible that $h_{1,2}$ decay into one another depending on their masses.
If $M_2 \le M_1/2$, it is kinematically allowed for $h_1$ to decay into a pair of $h_2$, 
which would increase the total decay width of $h_1$. 
However, in this case, we found that the total decay width of $h_1$ exceeds too much
the currently known value of the Higgs decay width 
$\Gamma_\textrm{exp}= 3.7^{+1.9}_{-1.4}$ MeV \cite{ParticleDataGroup:2024cfk}. 
Therefore, we only consider the case of heavy scalar boson $h_2$ with mass $M_2 > M_1/2$
as similarly done in Ref.~\cite{Kim:2018ecv}. 
Along with creating new interactions between scalar bosons, 
the Higgs portal terms also modify the Higgs self-couplings substantially.
For instance, the deviation of experimental value of the Higgs triple coupling for $h_1^3$ interaction
from the SM expectation for $\sin\theta = 0.1$ lies within the expected precision of the VLHC experiment, 
but not within the HL-LHC precision.
Further detailed discussion on this can also be found in Ref.~\cite{Kim:2018ecv}. 

If $M_\phi < M_1/2$, similarly to the Type I case, 
the SM Higgs $h_1$ can decay invisibly into a pair of DM through mixing with decay width,
\beq
\Gamma\left(h_1 \to \phi_i\phi_i\right)
= \frac{(\lamhp+\lamsp \tth \tbt)^2 v_h^2}{32\pi (1+\tth^2)M_h}\sqrt{1-\frac{4M_\phi^2}{M_h^2}}. 
\eeq
In Fig. \ref{fig:sigmaSI2}, 
we plot the SI DM-nucleon scattering cross section 
by varying the DM mass $M_\phi$ using parameter sets allowed by the relic density observation
over the following parameter space,
\beq
\lambda_{h\phi} \in \left[0,~2\pi\right], \
M_{\phi}/{\rm GeV} \in \left[10,~10^5\right], \
M_2/{\rm GeV} \in \left[63,~800\right], \
\tbt \in \left[1,~10\right], \
\tth \in \left[0,~0.25\right]. 
\eeq
Our choice of the mixing angle $\tth \leq 0.25$ is quite safe against the LEP2 constraints 
since the $h_2$ mass exceeds the corresponding LEP2 lower mass bound.
There are additional experimental constraints on the DM annihilation cross section
from measurements by Fermi-LAT \cite{Fermi-LAT:2015att,Fermi-LAT:2017opo} and H.E.S.S.~\cite{HESS:2022ygk}.
The Fermi-LAT results set stringent limits on the annihilation cross sections for DM masses below a few hundred GeV.
For H.E.S.S., the limits are applied to DM masses in the range of 300~GeV to 7~TeV.
These constraints do not affect our results, 
as the estimated annihilation cross section in our model remains well below these limits.
Additionally, Big Bang nucleosynthesis constrains the lifetime of the additional scalar $h_2$ to be less than 1 s \cite{Jedamzik:2009uy,Kaplinghat:2013yxa}, 
which is consistent with our model.
The constraints from searches for a heavy Higgs at the LHC 
\cite{ATLAS:2022fpx, ATLAS:2023wqy, CMS:2021klu}
also do not impose significant restrictions on our analysis.

We focus on the $N = 2$ case only for simplicity, 
which corresponds to a scenario with two exact copies of the DM.
The DM-nucleon scattering in this model 
occurs through the two $t$-channel diagrams exchanging $h_1$ and $h_2$.
Due to the mixing between $h_1$ and $h_2$,
the SI cross section values of the allowed data are widely spread as shown in the figure.
As a result, any DM mass in the range 55 GeV $\lesssim M_\phi \lesssim 5.9$ TeV is allowed for $N = 2$.
In Fig. \ref{fig:lamhp_vs_Mdm2}, 
it is also shown that any value of the coupling $\lamhp$ is possible.
Similarly, the obtained results do not constrain the mixing angle $\theta$ as well.
This provides us with richer phenomenology and the potential for a stable vacuum, 
as we will discuss in the next section.

\section{Vacuum Stability and Perturbativity}

\subsection{Type I}

The scalar couplings $\lambda_{i}$ can grow significantly with increasing renormalization scale $\Lambda$,
and one can constraint those by applying the tree-level perturbative unitarity to scalar elastic scattering processes
for the zeroth partial wave amplitude \cite{Lee:1977eg,Marciano:1989ns,Cynolter:2004cq}.
The bounds on the couplings $\lambda_{h\phi,\phi}$ are given by
$|\lambda_{h\phi}| \leq 8\pi$ and $\lambda_{\phi} \leq 4\pi/3$.
The $\beta$-function of a coupling $\lambda_i$ at a scale $\mu$ in the RGE is defined as
$\beta_{\lambda_i} = \partial \lambda_i/\partial \log\mu$.
For dimensionless couplings in the scalar potential (including the DM Yukawa coupling),
the one-loop $\beta$-functions are given by 
\beqr \label{eq:betafunction}
\beta_{\lambda_{t}}^{(1)} &=& \frac{\lambda_{t}}{16\pi^2}\left[ \frac{9}{2}\lambda_t^2 
-\left(8g_3^2 + \frac{9}{4}g_2^2 + \frac{17}{12}g_1^2\right) \right],
\nnb \\[1pt]
\beta_{\lambda_h}^{(1)} &=& \frac{1}{16\pi^2}\left[ 24\lambda_h^2 + 12\lambda_h\lambda_t^2 -6\lambda_t^4
-3\lambda_h(3g_2^2 + g_1^2) +\frac{3}{8}\big(2g_2^4+(g_2^2+g_1^2)^2\big) + \frac{N}{2}\lambda_{h\phi}^2 \right],
\nnb \\[1pt]
\beta_{\lambda_{h\phi}}^{(1)} &=& \frac{\lambda_{h\phi}}{16\pi^2}\left[ 12\lambda_h 
-\frac{3}{2}(3g_2^2 + g_1^2) + 6\lambda_t^2 + 4\lambda_{h\phi} + 2(N+2)\lambda_\phi \right],
\nnb \\[1pt]
\beta_{\lambda_{\phi}}^{(1)} &=& \frac{1}{16\pi^2}\left[2\lambda_{h\phi}^2 + 2(N+8) \lambda_\phi^2 \right], 
\eeqr 
where $g_1$ and $g_2$ are the SM U$(1)_Y$ and SU$(2)_L$ couplings, respectively,
and $\lambda_t$ is the top Yukawa coupling.
While we consider the NP effects on the effective potential and the $\beta$-functions at one-loop order,
in order to show how much the NP contribution is needed for stabilizing the scalar potential,
we include the following two-loop $\beta$-functions for the top-Yukawa and Higgs quartic self-couplings
as done in Ref.~\cite{Baek:2012uj} because those contributions are sizable:
\beqr
\beta_{\lambda_{t}}^{(2)} &\simeq& \frac{\lambda_{t}}{(16\pi^2)^2}
\bigg[-12\lambda_t^4-12\lambda_t^2\lambda_h+6\lambda_h^2
+\lambda_t^2\left(36g_3^2+\frac{225}{16}g_2^2+\frac{131}{16}g_1^2\right)
\nnb \\[1pt]
&& +g_3^2\left(9g_2^2+\frac{19}{9}g_1^2\right) -108g_3^4 -\frac{3}{4}g_2^2g_1^2
-\frac{23}{4}g_2^4 + \frac{1187}{216}g_1^4 \bigg],
\nnb \\[1pt]
\beta_{\lambda_h}^{(2)} &\simeq& \frac{1}{(16\pi^2)^2}
\bigg[ \lambda_h\lambda_t^2\left(-144\lambda_h-3\lambda_t^2 
+ 80g_3^2 + \frac{85}{6}g_1^2 + \frac{45}{2}g_2^2 \right)
\nnb \\[1pt]
&& + \lambda_h\left(-312\lambda_h^2 + \lambda_h\left(36g_1^2 + 108g_2^2\right)
+\frac{629}{24}g_1^4 - \frac{73}{8}g_2^4 +\frac{39}{4}g_1^2g_2^2 \right)
\nnb \\[1pt]
&& + \lambda_t^2\left( 30\lambda_t^4 -\lambda_t^2\left(\frac{8}{3}g_1^2 + 32g_3^2\right) 
-\frac{19}{4}g_1^4 - \frac{9}{4}g_2^4 + \frac{21}{2}g_1^2g_2^2 \right) 
\nnb \\[1pt]
&& + \frac{1}{48}\left(915g_2^6 - 379g_1^6 - 289g_1^2g_2^4 - 559g_1^4g_2^2 \right) 
\bigg].
\eeqr
The $\beta$-functions for the SM gauge couplings are not altered by the NP couplings up to the next-leading order 
and can be found in Ref.~\cite{Schrempp:1996fb}.
For numerical simulation, we assume the central values for the top and gauge boson masses
and use the following SM values:
$g_1(M_t) = 0.359,\, g_2(M_t) = 0.648,\, g_3(M_t) = 1.167,\, \lambda_t (M_t) = 0.951,\, \lambda_h (M_t) = 0.129$. 
As one can see from Eq.~(\ref{eq:betafunction}),
the SM-DM interaction coupling $\lambda_{h\phi}$ give positive contributions to the $\beta$-function of the Higgs quartic.
Especially for large $N$, $\lambda_{h\phi}$ contributions are much enhanced.
Also, the contribution of $\lambda_{\phi}$ to the $\beta$-function of $\lambda_{h\phi}$ is sizable,
so the DM self-coupling is also important in stabilizing the scalar potential.

\begin{figure}[!ht]
\centering%
\subfigure[$\lambda_{h\phi}(v_h) = 0.6$]
{\label{fig:rge1} %
\includegraphics[width=7.4cm]{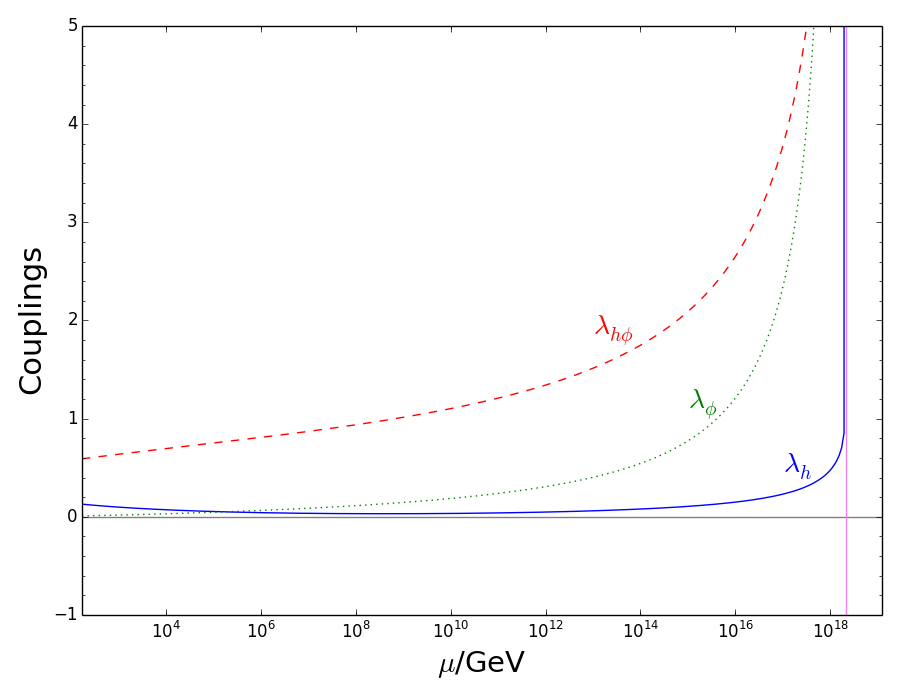}} \
\subfigure[$\lambda_{h\phi}(v_h) = 1.8$]
{\label{fig:rge2} %
\includegraphics[width=7.4cm]{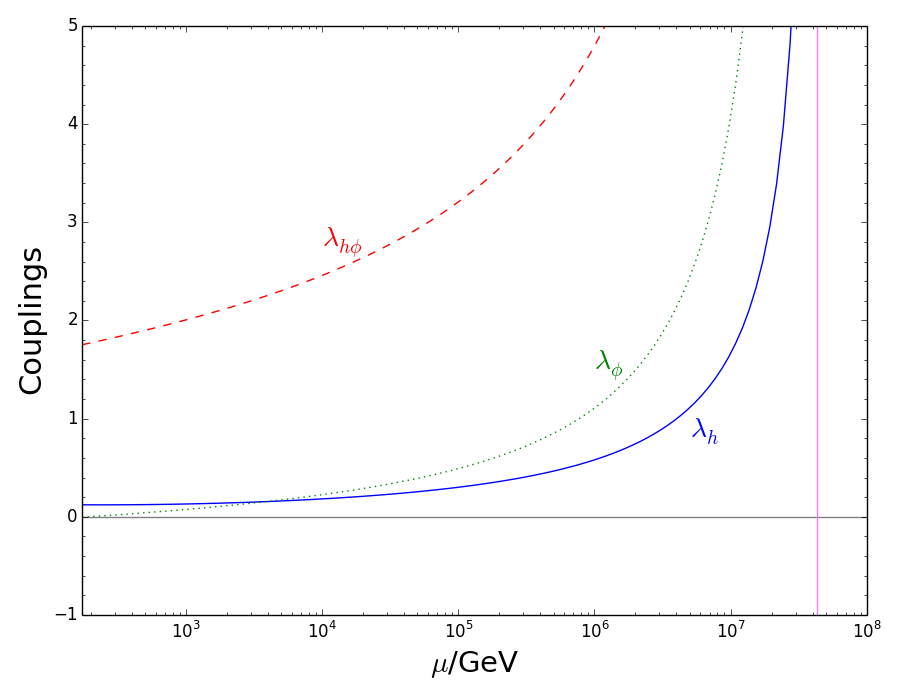}} 
\caption{Running of the massless couplings for (a) $\lambda_{h\phi}(v_h) = 0.6$ and (b) $\lambda_{h\phi}(v_h) = 1.8$ 
with the renormalization scale $\mu$ for $\lambda_{\phi}(v_h) = 0.01$.
The vertical (violet) line indicates where one of the couplings hit Landau pole.}
\label{fig:rgeI}
\end{figure}

As discussed in the previous section, 
the direct detection bounds severely constrain the DM mass and the coupling $\lamhp$.
In the low DM mass region, the allowed $\lamhp$ is very small, 
and the metastability of the Higgs vacuum remains unaffected.
In the high DM mass region, the lower bound of $\lamhp$ is approximately 0.6 for $N=1$ and 1.8 for $N=2$.
In Fig. \ref{fig:rgeI},
we plot the running of the massless couplings for $\lambda_{h\phi}(v_h) = 0.6$ and 1.8, 
using the renormalization scale $\mu$ with $\lambda_{\phi}(v_h) = 0.01$.
The scalar couplings increase drastically and reach the Landau pole near 
$\mu = 2.2 \times 10^{18}$ GeV for $\lambda_{h\phi}(v_h) = 0.6$
and $\mu = 4.4 \times 10^7$ GeV for $\lambda_{h\phi}(v_h) = 1.8$. 
Therefore, if the DM mass is heavier than a few TeV, 
this model can be considered as an effective theory valid up to those scales. 
As a reference, we scanned the full range of $\lambda_{h\phi}(\mu)$ 
and present the lower and upper bounds of the coupling where the electroweak vacuum is stable in Table.~1.
If $\lambda_{h\phi}(v_h) \leq \lambda_1$, the electroweak vacuum is metastable.
Conversely, if $\lambda_2 \leq \lambda_{h\phi}(v_h)$, one of the couplings becomes non-perturbative below the Planck scale.
If there are other scalar couplings, such as those in the Type II model,
the boundary values above may change, but not drastically. 
Therefore, the result presented in Table~1 can serve as a guideline for similar models.

\begin{table}
\caption{
\label{tbl:ref}
The boundary values $\lambda_1$ and $\lambda_2$ of the region for $\lambda_{h\phi}(v_h)$
where the electroweak vacuum is stable for each value of $N$ and $\lambda_{\phi}(v_h) $.}
\begin{center}
\begin{tabular}{|>{\centering\arraybackslash}p{3em}
|>{\centering\arraybackslash}p{4em}
|>{\centering\arraybackslash}p{4em}
|>{\centering\arraybackslash}p{4em}|
}
\hline
\centering\multirow{2}{3em}{\centering $N$} &
\centering\multirow{2}{3em}{\centering $\lambda_{\phi}(v_h)$} & 
\multicolumn{2}{c|}{$\lambda_{h\phi}(v_h)$} \\
\cline{3-4}
&& $\lambda_1$ & $\lambda_2$ \\
\hline
\hline 
\centering\multirow{2}{3em}{\centering 1}& 
0.01 & $0.375$ & $0.572$ \\
\cline{2-4}
& 0.1 & $0.357$ & $0.474$ \\
\cline{2-4}
\hline
\hline 
\centering\multirow{2}{3em}{\centering 2}& 
0.01 & $ 0.278$ & $0.538$ \\
\cline{2-4}
& 0.1 & $0.260$ & $0.446$ \\
\cline{2-4}
\hline
\hline 
\centering\multirow{2}{3em}{\centering 4}& 
0.01 & $0.202 $ & $0.472$ \\
\cline{2-4}
& 0.1 & $0.184$ & $0.387$ \\
\cline{2-4}
\hline 
\end{tabular}
\end{center}
\end{table}

\subsection{Type II}

Due to the scalar mixing terms added by the scalar mediator,
the one-loop $\beta$-functions are extended as 
\beqr
\label{eq:TypeII-Beta}
\beta_{\lambda_h}^{(1)} &=& \frac{1}{16\pi^2}\left[ 24\lambda_h^2 + 12\lambda_h\lambda_t^2 -6\lambda_t^4
-3\lambda_h(3g_2^2 + g_1^2) +\frac{3}{8}\big(2g_2^4+(g_2^2+g_1^2)^2\big) + \half\lambda_{hs}^2 
+ \frac{N}{2}\lambda_{h\phi}^2 \right],
\nnb \\[1pt]
\beta_{\lambda_{hs}}^{(1)} &=& \frac{1}{16\pi^2}\left[ \lambda_{hs}\left(12\lambda_h 
-\frac{3}{2}(3g_2^2 + g_1^2) + 6\lambda_t^2 + 4\lambda_{hs} + 6\lambda_s\right) 
+ N\lambda_{h\phi}\lambda_{s\phi} \right],
\nnb \\[1pt]
\beta_{\lambda_s}^{(1)} &=& \frac{1}{16\pi^2}\left[2\lambda_{hs}^2 + 18 \lambda_s^2 
+ \frac{N}{2}\lambda_{s\phi}^2 \right], 
\nnb \\[1pt]
\beta_{\lambda_{h\phi}}^{(1)} &=& \frac{1}{16\pi^2}\left[ \lambda_{h\phi}\left(12\lambda_h 
-\frac{3}{2}(3g_2^2 + g_1^2) + 6\lambda_t^2 + 4\lambda_{h\phi} + 2(N+2)\lambda_\phi\right) 
+ \lambda_{hs}\lambda_{s\phi} \right],
\nnb \\[1pt]
\beta_{\lambda_{s\phi}}^{(1)} &=& \frac{1}{16\pi^2}\bigg[\lambda_{s\phi}\bigg(4\lambda_{s\phi}
+ 6\lambda_s + 2(N+2)\lambda_\phi \bigg) + 4\lambda_{hs}\lambda_{h\phi} \bigg], 
\nnb \\[1pt]
\beta_{\lambda_{\phi}}^{(1)} &=& \frac{1}{16\pi^2}\left[2\lambda_{h\phi}^2 + 2(N+8) \lambda_\phi^2 
+ \half \lambda_{s\phi}^2 \right].
\eeqr 
The perturbative unitarity bounds on the additional scalar couplings are given similarly as
$|\lambda_{s\phi, hs}| \leq 8\pi$ and $\lambda_{s} \leq 4\pi/3$.

\begin{figure}[!ht]
\centering%
\includegraphics[width=7.4cm]{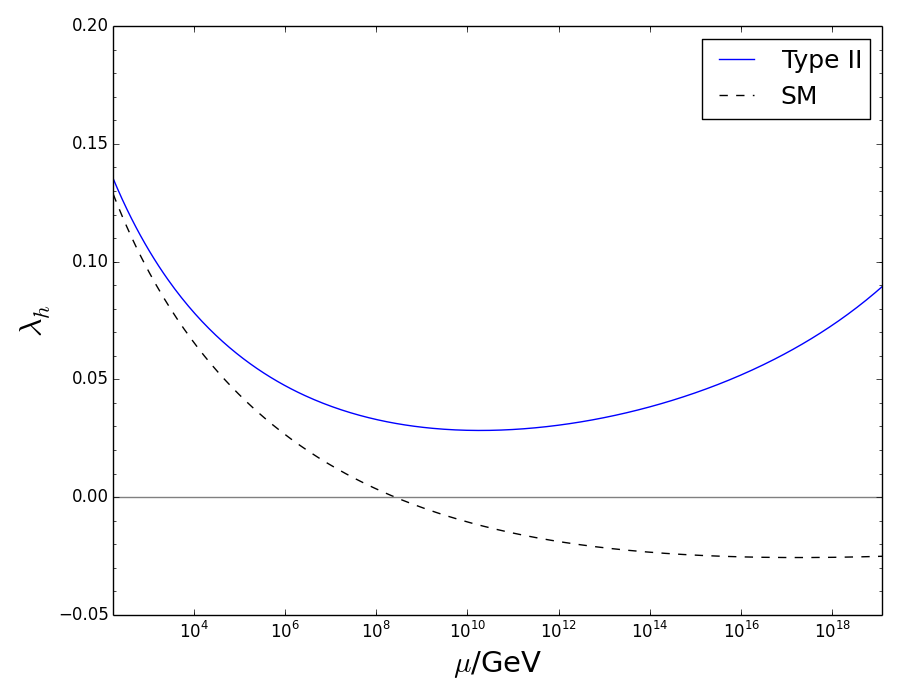} \
\includegraphics[width=7.4cm]{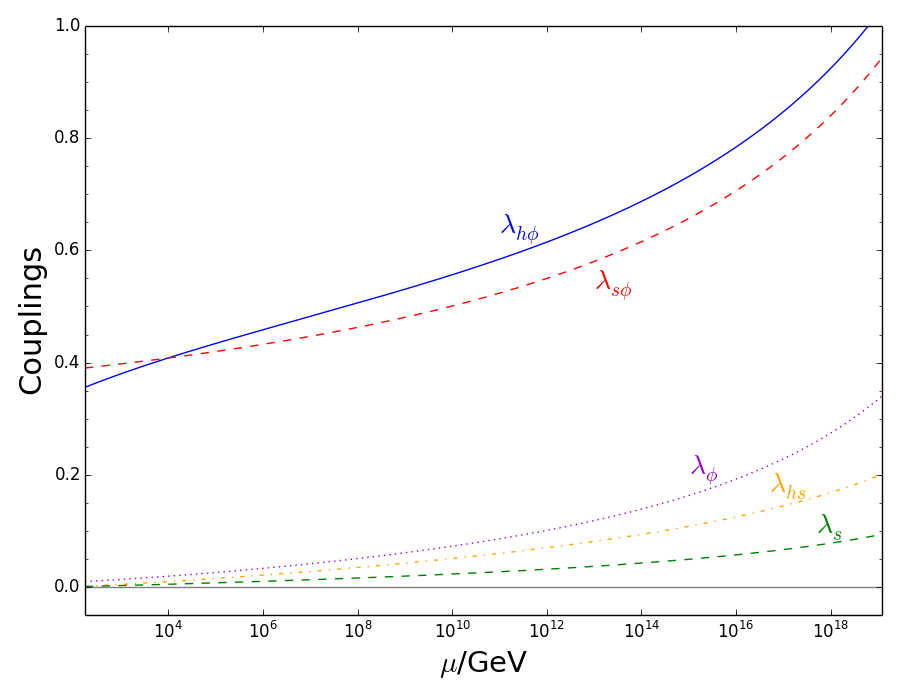} 
\caption{Running of the massless couplings given in Eq.~(\ref{eq:bench})
with the renormalization scale $\mu$ for $N = 2$.
}
\label{fig:rgeII}
\end{figure}

As a specific numerical example, we select a benchmark point 
(the red dot in Figs.~\ref{fig:sigmaSI} and \ref{fig:lamhp_vs_Mdm}),
which corresponds to the following model parameters:
\beqr
\label{eq:bench}
&& \tth = 0.0918, \quad M_2 = 83.4\, \textrm{GeV}, \quad M_\phi = 919.6\, \textrm{GeV}, \quad \tbt = 7.97,
\nnb \\[1pt]
&& \lamh = 0.128, \quad \lamhs = 0.0017, \quad \lams = 0.00085, 
\nnb \\[1pt]
&& \lamhp = 0.361, \quad \lamsp = 0.392, \quad \lamp = 0.01.
\eeqr 
In Fig.~\ref{fig:rgeII}, 
we plot the running behavior of the dimensionless scalar couplings for the case of $N=2$,
where the scalar vacua are stable and all scalar couplings remain perturbative (less than $4\pi$) up to the Planck scale. 
It is evident that the nonzero coupling $\lamhp$ plays a crucial role in stabilizing the Higgs potential.
In the high mass region of the DM $\phi$, 
either of the phenomenologically allowed quadratic couplings, $\lamhp$ and $\lamsp$, are large enough 
to be non-perturbative at high-energy scales.
We found that within the parameter sets that satisfy the vacuum stability and perturbativity conditions, 
the DM mass is typically around the 1 TeV mark.
If one chooses a larger DM self-coupling $\lamp$, 
the boundary values of the region for $\lambda_{h\phi}$, where the electroweak vacuum remains stable,
decrease as shown in Table 1. 
As such, the preferred value of the DM mass based on the vacuum stability condition may vary, but not significantly.
The DM self-coupling $\lamp$ can be constrained by DM self-interactions in small-scale galaxies. 
However, a comprehensive study that varies all parameters is beyond the scope of this paper.

\section{Conclusion}
In this work, we studied two models involving a scalar dark matter multiplet $\phi$ of a global $O(N)$ symmetry.
The type I model includes only the scalar DM multiplet, which is connected to the Higgs sector in the SM.
In comparison, the Type II model extends the hidden sector by introducing a new scalar mediator $S$ 
governed by $Z_{2}$ symmetry.

For the experimental constraints on these models,
we considered the branching ratio of the Higgs invisible decay, the relic abundance of DM, 
and the SI DM-nucleon scattering cross sections measured at XENON1T, PandaX4T, and LZ.
These constraints are more stringent than the naive upper bound on the DM mass $M_{\phi}$
derived from the partial-wave unitarity of the $S$-matrix.
The allowed regions for $M_{\phi}$ versus the Higgs-DM coupling $\lambda_{h\phi}$
for both models are shown in Fig.~\ref{fig:lamhp_vs_Mdm}.
For the Type I model, the allowed regions for $M_{\phi}$ are near the resonance $M_{\phi}\sim M_{h}/2$
and in the high mass region above a few TeV.
In the high mass region, the lower bounds on $M_{\phi}$ increase as $N$ grows.
When $N\geq 11$, the Type I model is excluded by the LZ bound in the high mass region.
In Type II, the allowed regions for $M_{\phi}$ and $\lambda_{h\phi}$ are
broadened due to mixing between the SM Higgs $h_1$ and the new scalar $h_2$.

Regarding the allowed model parameters, we investigated the vacuum stability of the Higgs potential 
and the perturbative behavior of the new scalar quartic and quadratic couplings. 
As shown in the $\beta$-functions for $\lambda_{h}$ in Eqs.~\eqref{eq:betafunction} and \eqref{eq:TypeII-Beta},
nonzero $\lambda_{h\phi}$ is crucial in stabilizing the Higgs potential.
In Table~\ref{tbl:ref},
we presented reference values for the lower and upper bounds of $\lambda_{h\phi}$ required for stabilizing the electroweak vacuum.
It is important to note that the Type I model cannot achieve a stable vacuum, while the Type II model can. 
For the Type II model, we illustrated the running behavior of the scalar couplings at a benchmark point 
where the electroweak vacuum remains stable and all couplings stay perturbative up to the Planck scale in Fig.~\ref{fig:rgeII}.
Our result can be applied to study similar new physics models involving the scalar DM connected to the SM 
through the Higgs portal with or without an additional scalar mediator.

\section*{Acknowledgments}
This work is supported by Basic Science Research Program through the National Research Foundation of Korea (NRF) 
funded by the Ministry of Education under the Grant No. RS-2023-00248860 (S.-h.~Nam)
and also funded by the Ministry of Science and ICT under the Grants No. NRF-2020R1A2C3009918 (S.-h.~Nam and J.~Lee).


\begin{thebibliography}{99}
\bibitem{Silveira:1985rk} V.~Silveira and A.~Zee, 
Phys. Lett. B \textbf{161}, 136-140 (1985). 

\bibitem{McDonald:1993ex} J.~McDonald, 
Phys. Rev. D \textbf{50}, 3637-3649 (1994).

\bibitem{Holz:2001cb} D.~E.~Holz and A.~Zee, 
Phys. Lett. B \textbf{517}, 239-242 (2001). 

\bibitem{Burgess:2000yq} C.~P.~Burgess, M.~Pospelov and T.~ter Veldhuis, 
Nucl. Phys. B \textbf{619}, 709-728 (2001). 

\bibitem{He:2007tt} X.~G.~He, T.~Li, X.~Q.~Li and H.~C.~Tsai, 
Mod. Phys. Lett. A \textbf{22}, 2121-2129 (2007). 

\bibitem{He:2011de} X.~G.~He and J.~Tandean, 
Phys. Rev. D \textbf{84}, 075018 (2011).

\bibitem{Drozd:2011pex}
A.~Drozd, B.~Grzadkowski and J.~Wudka,
Acta Phys. Polon. B \textbf{42}, no.11, 2255 (2011).


\bibitem{Isidori:2001bm}
G.~Isidori, G.~Ridolfi and A.~Strumia,
Nucl. Phys. B \textbf{609}, 387-409 (2001).

\bibitem{Degrassi:2012ry}
G.~Degrassi, S.~Di Vita, J.~Elias-Miro, J.~R.~Espinosa, G.~F.~Giudice, G.~Isidori and A.~Strumia,
JHEP \textbf{08}, 098 (2012).

\bibitem{Alekhin:2012py}
S.~Alekhin, A.~Djouadi and S.~Moch,
Phys. Lett. B \textbf{716}, 214-219 (2012).

\bibitem{Buttazzo:2013uya}
D.~Buttazzo, G.~Degrassi, P.~P.~Giardino, G.~F.~Giudice, F.~Sala, A.~Salvio and A.~Strumia,
JHEP \textbf{12}, 089 (2013).

\bibitem{Bednyakov:2015sca}
A.~V.~Bednyakov, B.~A.~Kniehl, A.~F.~Pikelner and O.~L.~Veretin,
Phys. Rev. Lett. \textbf{115}, no.20, 201802 (2015).

\bibitem{Gonderinger:2009jp}
M.~Gonderinger, Y.~Li, H.~Patel and M.~J.~Ramsey-Musolf,
JHEP \textbf{01}, 053 (2010).

\bibitem{Lebedev:2012zw}
O.~Lebedev,
Eur. Phys. J. C \textbf{72}, 2058 (2012).


\bibitem{Bento:2000ah} M.~C.~Bento, O.~Bertolami, R.~Rosenfeld and L.~Teodoro, 
Phys. Rev. D \textbf{62}, 041302 (2000). 

\bibitem{McDonald:2001vt} J.~McDonald, 
Phys. Rev. Lett. \textbf{88}, 091304 (2002). 

\bibitem{Bernal:2015xba} N.~Bernal and X.~Chu, 
JCAP \textbf{01}, 006 (2016).


\bibitem{Tulin:2017ara} S.~Tulin and H.~B.~Yu, 
Phys. Rept. \textbf{730}, 1-57 (2018).


\bibitem{Kim:2022sfc}
Y.~G.~Kim, K.~Y.~Lee, J.~Lee and S.~h.~Nam,
Phys. Rev. D \textbf{106}, no.9, 095004 (2022).

\bibitem{Jung:2019dog}
D.~W.~Jung, J.~Lee and S.~H.~Nam,
Phys. Lett. B \textbf{797}, 134823 (2019).

\bibitem{Claude:2021sye}
J.~Claude and S.~Godfrey,
Eur. Phys. J. C \textbf{81}, no.5, 405 (2021).


\bibitem{LEPWorkingGroupforHiggsbosonsearches:2003ing}
R.~Barate \textit{et al.} [LEP Working Group for Higgs boson searches, ALEPH, DELPHI, L3 and OPAL],
Phys. Lett. B \textbf{565}, 61-75 (2003).

\bibitem{LHCHiggsCrossSectionWorkingGroup:2016ypw}
D.~de Florian \textit{et al.} [LHC Higgs Cross Section Working Group],
arXiv:1610.07922 [hep-ph].

\bibitem{ATLAS:2023tkt}
G.~Aad \textit{et al.} [ATLAS],
Phys. Lett. B \textbf{842}, 137963 (2023).

\bibitem{Griest:1989wd}
K.~Griest and M.~Kamionkowski,
Phys. Rev. Lett. \textbf{64}, 615 (1990).

\bibitem{Planck:2018vyg}
N.~Aghanim \textit{et al.} [Planck],
Astron. Astrophys. \textbf{641}, A6 (2020);
erratum: Astron. Astrophys. \textbf{652}, C4 (2021).


\bibitem{Belanger:2018ccd}
G.~B\'elanger, F.~Boudjema, A.~Goudelis, A.~Pukhov and B.~Zaldivar,
Comput. Phys. Commun. \textbf{231}, 173-186 (2018).

\bibitem{Belyaev:2012qa}
A.~Belyaev, N.~D.~Christensen and A.~Pukhov,
Comput. Phys. Commun. \textbf{184}, 1729-1769 (2013).


\bibitem{XENON:2018voc}
E.~Aprile \textit{et al.} [XENON],
Phys. Rev. Lett. \textbf{121}, no.11, 111302 (2018).

\bibitem{PandaX-4T:2021bab}
Y.~Meng \textit{et al.} [PandaX-4T],
Phys. Rev. Lett. \textbf{127}, no.26, 261802 (2021).

\bibitem{LZ:2022lsv}
J.~Aalbers \textit{et al.} [LZ],
Phys. Rev. Lett. \textbf{131}, no.4, 041002 (2023).

\bibitem{ParticleDataGroup:2024cfk}
S.~Navas \textit{et al.} [Particle Data Group],
Phys. Rev. D \textbf{110}, no.3, 030001 (2024).


\bibitem{Kim:2018ecv}
Y.~G.~Kim, K.~Y.~Lee and S.~H.~Nam,
Phys. Lett. B \textbf{782}, 316-323 (2018).

\bibitem{Fermi-LAT:2015att}
M.~Ackermann \textit{et al.} [Fermi-LAT],
Phys. Rev. Lett. \textbf{115}, no.23, 231301 (2015)

\bibitem{Fermi-LAT:2017opo}
M.~Ackermann \textit{et al.} [Fermi-LAT],
Astrophys. J. \textbf{840}, no.1, 43 (2017)

\bibitem{HESS:2022ygk}
H.~Abdalla \textit{et al.} [H.E.S.S.],
Phys. Rev. Lett. \textbf{129}, no.11, 111101 (2022)

\bibitem{Jedamzik:2009uy}
K.~Jedamzik and M.~Pospelov,
New J. Phys. \textbf{11}, 105028 (2009)

\bibitem{Kaplinghat:2013yxa}
M.~Kaplinghat, S.~Tulin and H.~B.~Yu,
Phys. Rev. D \textbf{89}, no.3, 035009 (2014)

\bibitem{ATLAS:2022fpx}
G.~Aad \textit{et al.} [ATLAS],
Eur. Phys. J. C \textbf{83}, no.6, 519 (2023)

\bibitem{ATLAS:2023wqy}
G.~Aad \textit{et al.} [ATLAS],
Phys. Lett. B \textbf{848}, 138394 (2024)

\bibitem{CMS:2021klu}
A.~Tumasyan \textit{et al.} [CMS],
Phys. Rev. D \textbf{105}, no.3, 032008 (2022)


\bibitem{Lee:1977eg}
B.~W.~Lee, C.~Quigg and H.~B.~Thacker,
Phys. Rev. D \textbf{16}, 1519 (1977).

\bibitem{Marciano:1989ns}
W.~J.~Marciano, G.~Valencia and S.~Willenbrock,
Phys. Rev. D \textbf{40}, 1725 (1989).

\bibitem{Cynolter:2004cq}
G.~Cynolter, E.~Lendvai and G.~Pocsik,
Acta Phys. Polon. B \textbf{36}, 827-832 (2005).

\bibitem{Baek:2012uj}
S.~Baek, P.~Ko, W.~I.~Park and E.~Senaha,
JHEP \textbf{11}, 116 (2012).

\bibitem{Schrempp:1996fb}
B.~Schrempp and M.~Wimmer,
Prog. Part. Nucl. Phys. \textbf{37}, 1-90 (1996).


\end{thebibliography}
\end{document}